# Direct Visualization of Gigahertz Acoustic Wave Propagation in Suspended Phononic Circuits

Daehun Lee[†,1], Qiyu Liu[†,2], Lu Zheng[1], Xuejian Ma[1], Mo Li[*,2,3], Keji Lai[*,1]

[1] Department of Physics, University of Texas at Austin, Austin TX 78712, USA

[2] Department of Electrical and Computer Engineering, University of Washington, Seattle, WA, 98195, USA

[3] Department of Physics, University of Washington, Seattle, WA, 98195, USA

[†] These authors contributed equally to this work
[*] E-mails: moli96@uw.edu; kejilai@physics.utexas.edu

## Abstract

We report direct visualization of gigahertz-frequency Lamb waves propagation in aluminum nitride phononic circuits by transmission-mode microwave impedance microscopy (TMIM). Consistent with the finite-element modeling, the acoustic eigenmodes in both a horn-shaped coupler and a sub-wavelength waveguide are revealed in the TMIM images. Using fast Fourier transform filtering, we quantitatively analyze the acoustic loss of individual Lamb modes along the waveguide and the power coupling coefficient between the waveguide and the parabolic couplers. Our work provides insightful information on the propagation, mode conversion, and attenuation of acoustic waves in piezoelectric nanostructures, which is highly desirable for designing and optimizing phononic devices for microwave signal processing and quantum information transduction.

## I. Introduction

Acoustic waves in the radiofrequency (MHz to GHz) range propagate in solid structures with a speed of several km/s that is 5 orders of magnitude slower than speed of light. Therefore, transduction from electromagnetic waves to acoustic waves enables signal processing on a dramatically slower timescale and much reduced device dimensions. Because acoustic waves cannot propagate in vacuum, radiative crosstalk between signal channels in acoustic devices is also much lower than in electromagnetic devices. As a result, various types of acoustic devices, such as surface acoustic wave (SAW), bulk acoustic wave (BAW), and flexural plate wave (FPW) devices, are widely utilized as delay lines, filters, oscillators, convolvers in wireless communication applications [1-3], and mass, pressure, and flow sensors in sensing applications [4]. Recently, propagating acoustic waves are considered universal quantum interconnects between different solid-state qubit systems, e.g., defect centers and superconducting qubits [5-10], for two reasons. First, the quantum states of these systems are highly susceptible to mechanical deformation with high coupling coefficients [11-15]; second, the acoustic wave can propagate with very low loss and noise at low temperatures [16,17]. There is also a strong interest in achieving efficient transduction between optical and microwave photons mediated by acoustic modes in optomechanical systems through acousto-optic coupling [18]. Optical and microwave-frequency acoustic (or phononic) modes are confined in wavelength-scale structures and interact through efficient acousto-optic coupling [19-26]. However, direct conversion from optical photons to acoustic phonons is intrinsically low in energy efficiency because of their large disparity in frequency (~ 5 orders of magnitude), assuming the same wavelengths. In contrast, converting MHz to GHz photons to acoustic phonons can be achieved much more efficiently using electromechanical transducers on piezoelectric materials [27]. To achieve efficient photon-phonon coupling, it is critical to engineer phononic structures to effectively guide and couple acoustic phonons into optomechanical systems.

A major challenge for designing phononic systems is that the density of state of acoustic phonons is very high [28], and different polarization modes inherently couple with each other through geometric deformation. Due to the small acoustic wavelength, it is often computationally expensive to perform a full 3D finite-element simulation of phononic structures with realistic sizes. The mechanical properties of materials are also more susceptible to fabrication processes than their optical properties, making simulations less accurate. In order to complement mechanical simulations, experimental probing of the acoustic fields has become an important field of research in recent years. For instance, MHz surface displacement fields have been imaged by scanning laser reflectormetry [29,30] or interferometry [31,32], stroboscopic X-ray imaging [33,34], scanning electron microscopy (SEM) [35,36], scanning tunneling microscopy (STM) [37,38], and nonlinear acoustic force microscopy (AFM) [39]. However, none of these techniques can simultaneously achieve sub-100 nm spatial resolution and > 1 GHz operation frequency, which are crucial for wavelength-scale acousto-optic devices. Thus, a new method that allows nanoscale investigation of wave phenomena, such as interference, diffraction, and localization, of GHz acoustic waves is desirable for designing and optimizing efficient optomechanical systems.

In this work, we report the visualization of 3.44 GHz Lamb waves in suspended aluminum nitride (AlN) phononic waveguides by transmission-mode microwave impedance microscopy (TMIM) [40,41]. The imaging results vividly demonstrate the coupling from anti-symmetrical membrane modes to waveguide modes through a parabolic horn-shaped coupler. Using fast Fourier transform (FFT) filtering, we identify individual waveguide modes and analyze their propagation loss along the waveguide. Our work provided insightful information on the propagation, attenuation, and coupling of Lamb waves in phononic circuits, which cannot be obtained by traditional microwave network analysis.

## II. Device and Simulation

The suspended phononic circuits in this work are fabricated on *c*-axis polycrystalline AlN thin films (thickness $t = 330$ nm) grown by magnetron sputtering on $SiO_2$/Si wafers. The circuit consists of an acoustic waveguide with width $w = 1$ μm and length $L = 100$ μm, connected to two identical parabolic horn-shaped acoustic couplers with a length of $L_h = 100$ μm. The couplers are designed to focus acoustic waves to the waveguide from two interdigital transducers (IDTs), one used as a transmitter and one as a receiver. The waveguide and couplers are patterned by standard electron beam lithography (EBL) and plasma etching of AlN, using 240-nm-thick $SiO_2$ as the hard mask. The IDT fingers have an aperture width of $A = 20$ μm. They are fabricated with EBL and the deposition of 7 nm Cr and 100 nm Au, followed by a standard lift-off process. To reduce the acoustic loss due to internal reflection and destructive interference in the IDT region, we use the split-finger design with a period of 3 μm and 4 fingers per period (inset of Fig. 1a). The 3$^{rd}$ harmonic mode with λ = 1 μm excited by this IDT is the Lamb wave to be investigated below.

Another layer of 7 nm Cr / 300 nm Au is deposited to thicken the bonding pads. Finally, the 3 μm thermal $SiO_2$ underneath the AlN film is removed using a vapor HF etcher to release the device from the substrate.

Electrical characterization of the AlN phononic circuit is carried out with a vector network analyzer (VNA). The transmitter and receiver IDTs are connected to the VNA through microwave cables and a pair of RF probes. The measured reflection coefficient ($S_{11}$) spectrum (Fig. 1b) shows a resonance at $f = 3.44$ GHz, which corresponds to the excitation of the Lamb mode at $\lambda = 1$ μm. The transmission coefficient ($S_{21}$) spectrum (Fig. 1c) between the two IDTs shows a smaller peak at the same frequency, indicating transmission of the Lamb wave through the circuit.

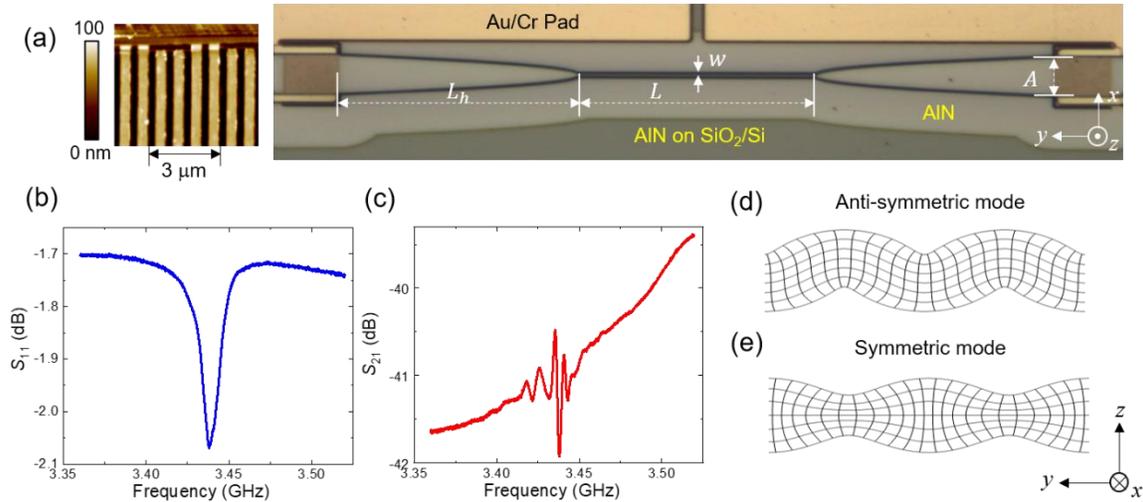

FIG. 1 **(a)** Optical image of the AlN phononic circuits. The inset on the left shows the AFM image of the split-finger IDT. **(b)** $S_{11}$ and **(c)** $S_{21}$ spectra of the device measured by a VNA, showing the acoustic resonance at $f = 3.44$ GHz. **(d)** Simulation results showing the anti-symmetric (A-mode) and **(e)** symmetric (S-mode) modes of the free-standing membrane at the IDT region.

To understand the excited Lamb mode, we model the IDT region as a large AlN membrane and simulate with the finite-element method (FEM). Figs. 1d and 1e show the the fundamental modes, where the displacement fields are either anti-symmetric (A-mode) or symmetric (S-mode) with respect to the *xy*-plane. Because the IDTs are patterned only on the top surface of the membrane, the electric fields generated by the IDTs are asymmetric along the thickness of the membrane. Consequently, the anti-symmetric membrane modes are predominantly excited. Therefore, the observed resonance at 3.44 GHz is attributed to the anti-symmetric Lamb mode, which agrees with the simulation results.

Compared with the membrane modes discussed above, which are only confined in the *z*-direction, the acoustic wave propagating in the suspended AlN circuits is also confined in the transverse *x*-direction. In the following, we will denote the Lamb modes in the subwavelength waveguide as waveguide modes. Fig. 2a shows the simulated dispersion relation of various modes on a waveguide with a width of 1 µm and a thickness of 330 nm. The dashed line denoted the excitation frequency at 3.44 GHz, which intersects with three branches of dispersion curves of the waveguide modes, hereafter labeled as S0, A1, and A0, respectively. Due to symmetry matching, the anti-symmetric membrane mode excited by the IDTs can only couple to the A0 and A1 modes. Fig. 2b and c display the simulated displacement fields of the two anti-symmetric modes (A0 and A1). Here the A0 mode in the lowest acoustic branch is the fundamental anti-symmetric breathing mode with $\lambda = 1$ µm, where the displacement field is uniformly distributed in the cross-section (*xz* plane). A1 with $\lambda = 2$ µm is the first-order anti-symmetric mode, where the displacement is out of phase between the center of the waveguide and the boundary of the waveguide. Detailed simulation results of the A0, S0, and A1 modes are included in Appendix A.

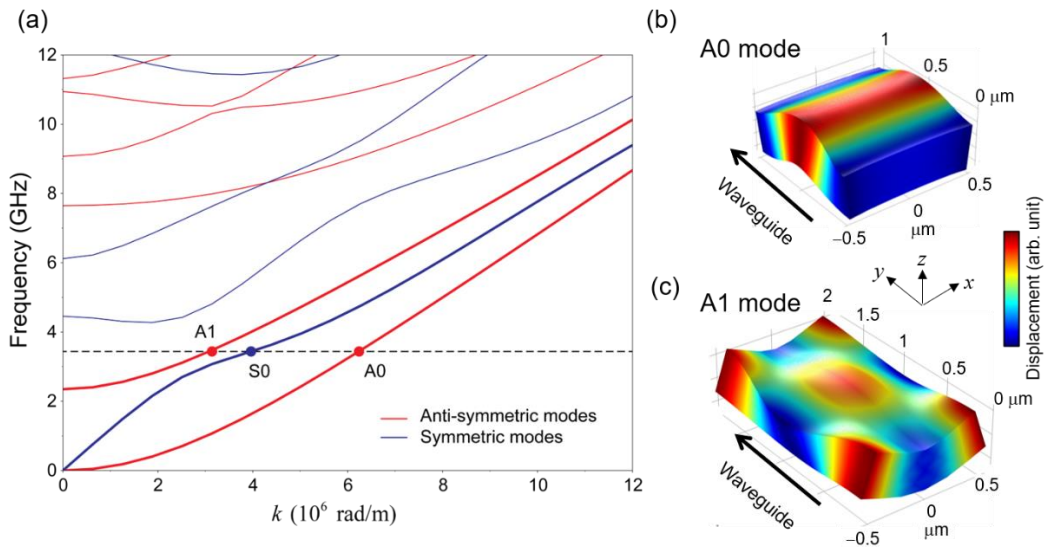

FIG. 2. **(a)** Calculated dispersion relation of a free-standing AlN phononic waveguide with 1 µm width. The red and blue curves represent anti-symmetric and symmetric modes, respectively. The intersection points between the dashed line at $f = 3.44$ GHz and three lowest branches of the dispersion curves are labeled as A0, S0, and A1 (see text). **(b)** 3D simulation results showing the mode shape of A0 and **(c)** A1 modes with color indicating the displacement magnitude.

## III. Experimental Results

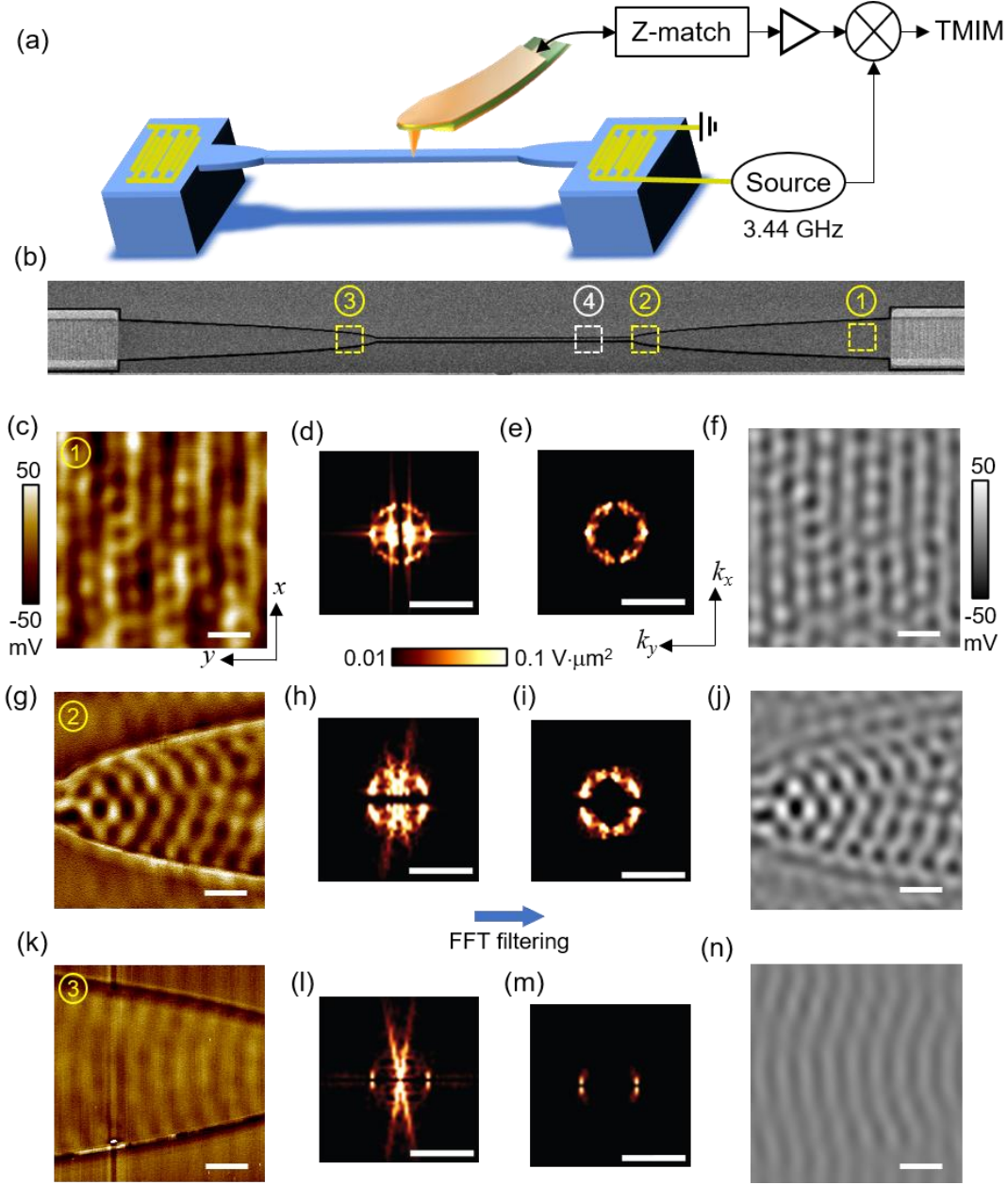

FIG. 3. **(a)** Schematics of the suspended photonic circiut and the TMIM setup. **(b)** SEM image of the device. The four dashed boxes show the locations where TMIM images are acquired. **(c)** TMIM image and **(d)** 2D FFT spectral image in Box #1. **(e)** FFT image after removing the diffusive spots at the center. **(f)** Inverse FFT image of (e). **(g – j)** Same as (c – e) in Box #2. **(k – n)** Same as (c – e) in Box #3. Scale bars are 2 μm for real-space images and $2\pi \cdot 2$ μm$^{-1}$ for $k$-space FFT images. The false-color scales for panels (c, g, k) are the same, so are (d, e, h, i, l, m), and (f, j, n).

Imaging of Lamb waves on the phononic circuit is carried out in our TMIM setup, an atomic-force-microscopy (AFM)-based technique with sub-100nm spatial resolution [40,41]. Fig. 3a shows the configuration of the TMIM experiment, where the acoustic wave is launched by the emitter IDT and the induced surface potential modulation at GHz is detected by the tip. The signal is then amplified and demodulated by an in-phase/quadrature (I/Q) mixer, using the same microwave source as the reference signal. The time-varying acoustic signal is thus converted to time-independent TMIM images, which are simultaneously acquired as the topographic image during the scanning (Appendix B). For simplicity, we will only present one of the two orthogonal TMIM channels in the following discussion.

Fig. 3b shows the SEM image of the AlN phononic circuit, where TMIM images were taken in several 10μm × 10μm areas marked by dashed boxes. Near the emitter IDT, the acoustic pattern in Fig. 3c is mostly planewave-like. The corresponding 2D FFT spectral image is shown in Fig. 3d. The large diffusive spots near the center of FFT data correspond to slow-varying background signals in the real space, presumably due to incoherent motion of the membrane. By filtering out this feature (Fig. 3e), one can see that the highest FFT intensity lies along the propagation direction ($y$ axis) with a wavevector $|k| = 2\pi \cdot 1$ μm$^{-1}$, consistent with $\lambda = 2\pi/k = 1$ μm of the anti-symmetric membrane mode. From the inverse FFT image in Fig. 3f, it is nevertheless obvious that Lamb waves with the same $|k|$ along other in-plane directions are present because of multiple reflections from boundaries of the parabolic coupler. At the bottom of this parabola (Fig. 3g), the wave front is strongly curved. Correspondingly, while the Lamb wave retains the wavevector $|k| = 2\pi \cdot 1$ μm$^{-1}$, the FFT intensity in the $y$-direction drops to zero (Fig. 3h). Finally, in the other parabolic coupler near the exit side of the waveguide, Lamb wave with the same $|k|$ but a much weaker amplitude is observed, as evidenced from both the raw and the FFT-filtered TMIM images (Fig. 3k – 3n).

We now focus our attention on acoustic modes in the suspended phononic waveguide. Fig. 4a shows the TMIM image taken in Box #4, where a complex waveform of the surface potential is clearly observed. By taking FFT (Fig. 4b) of the raw data, one can see that the TMIM results are a superposition of three distinct harmonic components. The fringe-like pattern along the $k_x$ direction in the FFT spectrum (Fig. 4d) corresponds to the double-slit feature in the TMIM image (Fig. 4c), which is from horizontal boundaries of the waveguide. Because of the finite size of the AFM tip, the waveguide appears to be slightly wider in the TMIM image than its actual width of 1 μm. Other than this topographic crosstalk, the most prominent FFT features are the four bright lines at $|k_y| = 2\pi \cdot 0.5$ μm$^{-1}$ (Fig. 4f). The inverse FFT image in Fig. 4e reveals that they are associated with the A1 mode with out-of-phase motion between the center and boundary of the waveguide, as depicted in Fig. 2c. Finally, the FFT image also display weak but discernible features at $|k_y| = 2\pi \cdot 1$ µm$^{-1}$ (Fig. 4h). The corresponding real-space image in Fig. 4g suggests that this is the A0 mode with in-phase particle motion across the width of the waveguide (Fig. 2b). The line profiles through the center of Figs. 4e and 4g are plotted in the corresponding insets, showing $\lambda = 2$ µm and 1 µm for the two modes, respectively. By comparing the amplitudes of the two modes, it is obvious that the A1 mode is the dominant mode excited in the waveguide. In other words, the anti-symmetric membrane mode in the parabolic coupler is mostly converted to the A1 mode in the waveguide, presumably due to mode conversion occurring near the tip of the parabolic coupler where the wavefront is distorted from the planar wavefront from the IDT (Fig. 3g-j). This conversion is undesired for many applications where the A0 mode in the waveguide is prefeered [26]. Therefore, a better design of the coupler will be needed and the TMIM measurement can provide critical insights.

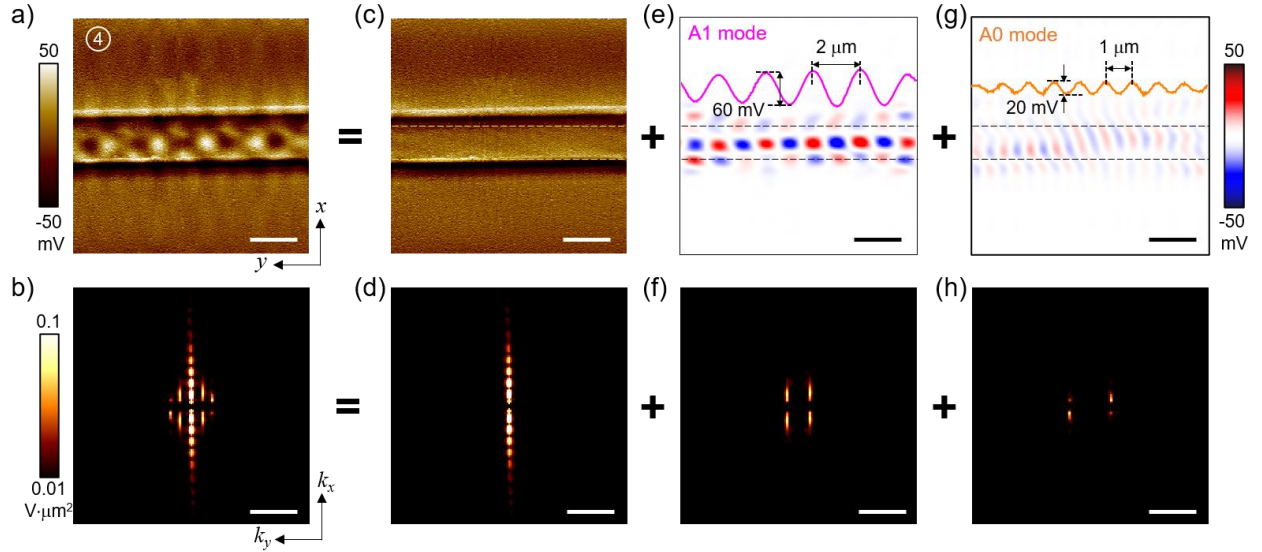

FIG. 4. **(a)** TMIM image and **(b)** its 2D FFT spectral image in Box #4 in Fig. 3b. **(c)** Filtered TMIM and **(d)** FFT images of the topographic artifact due to tranches on both sides of the waveguide. **(e)** Filtered TMIM and **(f)** FFT images associated with the A1 mode. **(g)** Filtered TMIM and **(h)** FFT images associated with the A0 mode. Insets of (e) and (g) show line cuts through center of the images. Note that (a) is the superposition of (c, e, g) in the real space and (b) the superposition of (d, f, h) in the $k$-space. Scale bars are 2 µm for real-space images and $2\pi \cdot 2$ µm$^{-1}$ for $k$-space FFT images.

The FFT filtering method described above allows us to remove the topographic crosstalk from the TMIM data and analyze oscillating amplitudes of individual waveguide modes. Figs. 5a and 5b plot the TMIM signals of A1 and A0 mode through the 100 µm long suspended waveguide, respectively. Representative FFT-filtered images are also shown near the entrance and exit points of the waveguide. For the primary A1 mode, the amplitude drops by a factor of ~ 1.5 over a length of 100 μm. The acoustic power loss is thus ~ 35 dB/mm, which is reasonable for the narrow waveguide. On the other hand, the decay of A0 mode is smaller ~ 20 dB/mm [26], although the error bar is large due to the weak signals and distorted wave profiles.

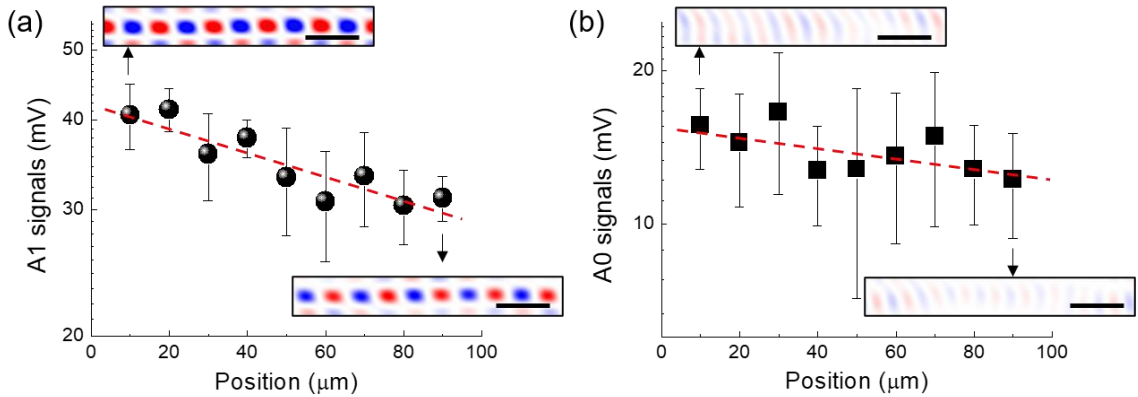

FIG. 5. **(a)** TMIM signals of the A1 and **(b)** A0 modes along the 100 μm waveguide. The insets show the filtered TMIM images associated with the A1 and A0 modes near the entrance and exit points of the waveguide. The red dashed lines are linear fits to the semi-log plots. All scale bars are 2 μm.

**IV. Discssions**

Quantitative analysis of the TMIM results reveals important information about the acoustic mode envolution in the suspended phononic circuit that cannot be obtained by two-port measurements. The $S_{21}$ data in Fig. 1c, for instance, convolve the piezoelectric and inverse piezoelectric transduction at the two IDTs, the acoustic propagation in the two parabolic couplers, and mode conversion in and out of the waveguide. In contrast, the TMIM imaging and FFT filtering allow us to focus on the waveguide modes and extract their acoustic loss in the narrow waveguide. The imaging method also reveals the acoustic mode coupling between the parabolic horn and the waveguide. When the Lamb wave enters the waveguide, the TMIM peak-to-peak signal drops from ~ 100 mV on the coupler side (Fig. 3g) to ~ 60 mV in the waveguide (Fig. 4e). Since the acoustic power is proportional to the square of oscillation amplitude and the width of the free-standing film, we can calculate a power coupling coefficient of ~ 20% by assuming a ratio of 2:1 in effective width near the entrance point. After the propagation of 100 μm, the TMIM signal drops to ~ 40 mV at the end of the waveguide. Using the same coupling coefficient, one can estimate a TMIM signal of 10 ~ 15 mV when entering the left parabolic coupler, consistent with the measured data in Fig. 3k. From the measured $S_{21}$ of ~ −40 dB, we obtain an electromechanical power conversion factor of 6 ~ 7% at the IDTs, which matches well with that of typical IDTs on AlN membranes [21,22,26]. As a result, the TMIM experiment provides a quantitative picture of various components in the phononic device down to the sub-wavelength scale.

Our work in visualizing GHz Lamb waves exemplifies the ability of TMIM to perform highly sensitive nanoscale acoustic imaging. From the calculated electromechanical power coupling efficiency, one can estimate the mechanical oscillation amplitude of 10 pm under an input power of 10 mW to the IDT [40]. Furthermore, the good signal-to-noise ratio in the TMIM images indicates that the detection limit at a normal scan rate of 10 min per frame is on the order of 0.1 pm. This level of surface acoustic vibration is extremely challenging for scanning laser interferometry [31,32], stroboscopic X-ray imaging [33,34], and nonlinear acoustic force microscopy [39]. More importantly, at the operation frequency of ~ 3 GHz, the acoustic wavelength of 1 ~ 2 µm is too small for optics-based techniques whose spatial resolution is diffraction limited. The AFM-based TMIM experiment, on the other hand, can routinely resolve sub-100nm features in the acoustic imaging. For even higher frequencies in the 10 GHz regime, which is of critical importance for optomechanics and quantum acoustics, TMIM might be the only technique of choice to map out the acoustic patterns on complex device structures [41].

## V. Conclusion

To summarize, we report the fabrication of suspended AlN acoustic waveguides and the visualization of 3.44 GHz Lamb waves on such phononic devices. Combining finite-element modeling and transmission-mode microwave microscopy, we are able to identify the membrane and waveguide modes and quantitatively analyze the acoustic coupling between the sub-wavelength waveguide and a pair of parabolic couplers. The FFT filtering allows us to separate the contribution from the two eigenmodes of the waveguide and calculate their acoustic loss. Our work demonstrates the exquisite sensitivity and high resolution of the TMIM technique, which is expected to find future applications in electromechanics, optomechanics, and quantum science and engineering.

**ACKNOWLEDGMENTS**

The TMIM work was supported by NSF Division of Materials Research Award DMR-2004536 and Welch Foundation Grant F-1814. The data analysis was partially supported by the NSF through the Center for Dynamics and Control of Materials, an NSF Materials Research Science and Engineering Center (MRSEC) under Cooperative Agreement DMR-1720595. The phononic waveguide work was supported by NSF Award EFMA-1741656 and EFMA-1641109. Part of this work was conducted at the Washington Nanofabrication Facility / Molecular Analysis Facility, a National Nanotechnology Coordinated Infrastructure (NNCI) site at the University of Washington with partial support from the National Science Foundation via awards NNCI-1542101 and NNCI-2025489.

## APPENDIX A: Finite-element modeling of the waveguide modes

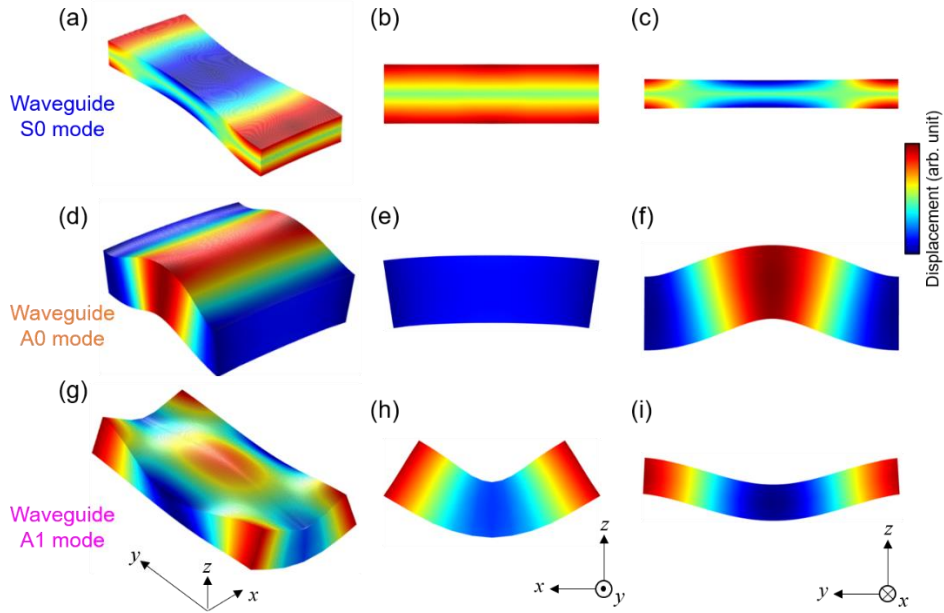

FIG. 6. Finite-element modeling results of the **(a-c)** S0, **(d-f)** A0, and **(g-i)** A1 modes the suspended AlN waveguide (330 nm in thickness and 1 μm in width). Panels (a, d, g) are the 3D views. Panels (b, e, h) are projections in the *xz*-plane. Panels (c, f, i) are projections in the *yz*-plane.

## APPENDIX B: AFM and TMIM images in a large field of view

Fig. 7 shows the simultaneously acquired AFM and TMIM images of the AlN waveguide device in a large field of view. The I/Q mixer in the TMIM electronics generates two orthogonal output channels, as displayed in Figs. 7b and 7c. The total TMIM signals, as plotted in Fig. 5, are vector sum of signals from the two channels. A plot of the TMIM-2 signal in Fig. 7d along the center of the waveguide shows the small decay of acoustic waves.

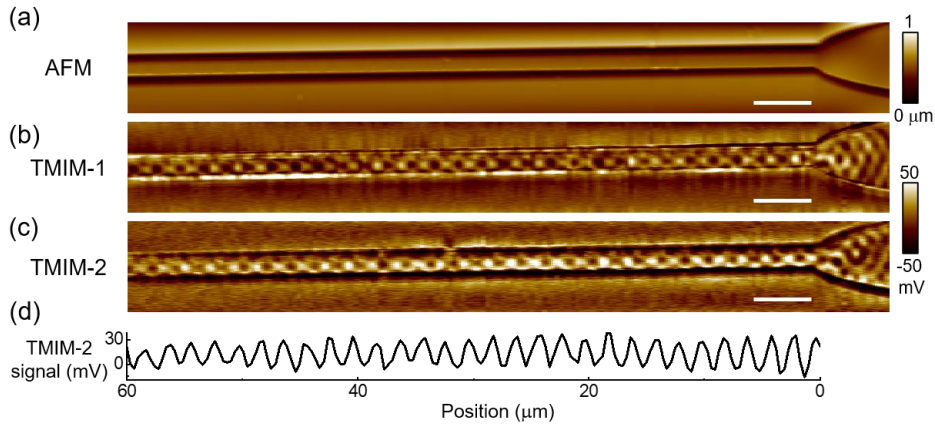

FIG. 7. **(a)** AFM, **(b)** TMIM-1, and **(c)** TMIM-2 images in a large field of view. Scale bars are 5μm. **(d)** TMIM-2 signals along the center of the waveguide in (c).

**References:**

[1] C. Campbell, *Surface acoustic wave devices for mobile and wireless communications* (Academic Press, San Diego, 1998), Applications of modern acoustics.
[2] K.-Y. Hashimoto, *Surface acoustic wave devices in telecommunications : modelling and simulation* (Springer, Berlin ; New York, 2000).
[3] D. P. Morgan, *Surface acoustic wave filters : with applications to electronic communications and signal processing* (Academic Press, Amsterdam ; London, 2007), 2nd edn.
[4] D. Ballantine Jr, R. M. White, S. J. Martin, A. J. Ricco, E. Zellers, G. Frye, and H. Wohltjen, *Acoustic wave sensors: theory, design and physico-chemical applications* (Elsevier, 1996).
[5] M. V. Gustafsson, T. Aref, A. F. Kockum, M. K. Ekstrom, G. Johansson, and P. Delsing, Science **346**, 207 (2014).
[6] M. J. A. Schuetz, E. M. Kessler, G. Giedke, L. M. K. Vandersypen, M. D. Lukin, and J. I. Cirac, Physical Review X **5**, 031031 (2015).
[7] Y. Chu, P. Kharel, W. H. Renninger, L. D. Burkhart, L. Frunzio, P. T. Rakich, and R. J. Schoelkopf, Science **358**, 199 (2017).
[8] A. Noguchi, R. Yamazaki, Y. Tabuchi, and Y. Nakamura, Phys. Rev. Lett. **119**, 180505 (2017).
[9] M. K. Ekström, T. Aref, J. Runeson, J. Björck, I. Boström, and P. Delsing, Appl. Phys. Lett. **110**, 073105 (2017).
[10] B. A. Moores, L. R. Sletten, J. J. Viennot, and K. W. Lehnert, Phys. Rev. Lett. **120**, 227701 (2018).
[11] E. R. MacQuarrie, T. A. Gosavi, N. R. Jungwirth, S. A. Bhave, and G. D. Fuchs, Phys. Rev. Lett. **111**, 227602 (2013).
[12] P. Ovartchaiyapong, K. W. Lee, B. A. Myers, and A. C. Jayich, Nat Commun **5**, 4429 (2014).
[13] D. A. Golter, T. Oo, M. Amezcua, K. A. Stewart, and H. Wang, Phys. Rev. Lett. **116**, 143602 (2016).
[14] K. W. Lee, D. Lee, P. Ovartchaiyapong, J. Minguzzi, J. R. Maze, and A. C. B. Jayich, Physical Review Applied **6**, 034005 (2016).
[15] S. Meesala *et al.*, Phys. Rev. B **97**, 205444 (2018).
[16] R. Manenti, M. J. Peterer, A. Nersisyan, E. B. Magnusson, A. Patterson, and P. J. Leek, Phys. Rev. B **93** (2016).
[17] E. B. Magnusson, B. H. Williams, R. Manenti, M. S. Nam, A. Nersisyan, M. J. Peterer, A. Ardavan, and P. J. Leek, Appl. Phys. Lett. **106**, 063509 (2015).
[18] M. Mirhosseini, A. Sipahigil, M. Kalaee, and O. Painter, Nature **588**, 599 (2020).
[19] F. M. Mayor, W. Jiang, C. J. Sarabalis, T. P. McKenna, J. D. Witmer, and A. H. Safavi-Naeini, Physical Review Applied **15**, 014039 (2021).
[20] W. Fu, Z. Shen, Y. Xu, C. L. Zou, R. Cheng, X. Han, and H. X. Tang, Nat Commun **10**, 2743 (2019).
[21] H. Li, Q. Liu, and M. Li, APL Photonics **4**, 080802 (2019).
[22] H. Li, S. A. Tadesse, Q. Liu, and M. Li, Optica **2**, 826 (2015).
[23] E. A. Kittlaus, H. Shin, and P. T. Rakich, Nat. Photon. **10**, 463 (2016).
[24] H. Shin, W. Qiu, R. Jarecki, J. A. Cox, R. H. Olsson, 3rd, A. Starbuck, Z. Wang, and P. T. Rakich, Nat Commun **4**, 1944 (2013).
[25] A. H. Safavi-Naeini, D. Van Thourhout, R. Baets, and R. Van Laer, Optica **6**, 213 (2019).
[26] Q. Y. Liu, H. Li, and M. Li, Optica **6**, 778 (2019).
[27] K. Yamanouchi and Y. Satoh, Japanese Journal of Applied Physics Part 1-Regular Papers Brief Communications & Review Papers **44**, 4532 (2005).
[28] M. S. Kushwaha, P. Halevi, L. Dobrzynski, and B. Djafari-Rouhani, Phys. Rev. Lett. **71**, 2022 (1993).
[29] A. Mahjoubfar, K. Goda, A. Ayazi, A. Fard, S. H. Kim, and B. Jalali, Appl. Phys. Lett. **98**, 101107 (2011).

[30] Y. Xu, W. Fu, C.-l. Zou, Z. Shen, and H. X. Tang, Appl. Phys. Lett. **112**, 073505 (2018).
[31] Y. Sugawara, O. Wright, O. Matsuda, M. Takigahira, Y. Tanaka, S. Tamura, and V. Gusev, Phys. Rev. Lett. **88**, 185504 (2002).
[32] D. M. Profunser, O. B. Wright, and O. Matsuda, Phys. Rev. Lett. **97**, 055502 (2006).
[33] R. Whatmore, P. Goddard, B. Tanner, and G. Clark, Nature **299**, 44 (1982).
[34] E. Zolotoyabko, D. Shilo, W. Sauer, E. Pernot, and J. Baruchel, Appl. Phys. Lett. **73**, 2278 (1998).
[35] D. Roshchupkin, T. Fournier, M. Brunel, O. Plotitsyna, and N. Sorokin, Appl. Phys. Lett. **60**, 2330 (1992).
[36] D. Roshchupkin, M. Brunel, R. Tucoulou, E. Bigler, and N. Sorokin, Appl. Phys. Lett. **64**, 164 (1994).
[37] W. Rohrbeck, E. Chilla, H.-J. Fröhlich, and J. Riedel, Appl. Phys. A **52**, 344 (1991).
[38] E. Chilla, W. Rohrbeck, H. J. Fröhlich, R. Koch, and K. Rieder, Appl. Phys. Lett. **61**, 3107 (1992).
[39] T. Hesjedal, Rep. Prog. Phys. **73**, 016102 (2009).
[40] L. Zheng, D. Wu, X. Wu, and K. Lai, Physical Review Applied **9**, 061002 (2018).
[41] L. Zheng, L. Shao, M. Loncar, and K. Lai, IEEE Microwave Magazine **21**, 60 (2020).